# Affective State Detection using fNIRs and Machine Learning

Ritam Ghosh
*Dept. of Electrical Engineering*
*Vanderbilt University*
Nashville, USA
ritam.ghosh@Vanderbilt.Edu

*Abstract* — **Affective states regulate our day to day to function and has a tremendous effect on mental and physical health. Detection of affective states is of utmost importance for mental health monitoring, smart entertainment selection and dynamic workload management. In this paper, we discussed relevant literature on affective state detection using physiology data, the benefits and limitations of different sensors and methods used for collecting physiology data, and our rationale for selecting functional near-infrared spectroscopy. We present the design of an experiment involving nine subjects to evoke the affective states of meditation, amusement and cognitive load and the results of the attempt to classify using machine learning. A mean accuracy of 83.04% was achieved in three class classification with an individual model; 84.39% accuracy was achieved for a group model and 60.57% accuracy was achieved for subject independent model using leave one out cross validation. It was found that prediction accuracy for cognitive load was higher (evoked using a pen and paper task) than the other two classes (evoked using computer bases tasks). To verify that this discrepancy was not due to motor skills involved in the pen and paper task, a second experiment was conducted using four participants and the results of that experiment has also been presented in the paper.**

*Keywords—affective states, fNIRs, brain imaging, BCI*

## I. INTRODUCTION

Affective states regulate our daily function and has a tremendous effect on our mental and physical health. A lot of research has been undertaken recently on detection and classification of affective states using automated techniques. Knowledge of affective states can be used in various applications like judging the mental health of a subject and evaluating the effectiveness of any therapy or intervention, particularly for people in the autism spectrum, who do not exhibit similar facial expressions as neurotypical individuals [1,2]. Affective states like amusement or sorrow can provide insights into an individual's mental state. Quantitative measurement of stress or cognitive overload are used for various applications like adaptive difficulty control of games or other physical activities [3,4], dynamic workload management to ensure optimum productivity and preventing overburdening of employees in the workplace [5,6]. An accurate and automated real time measurement of affective states can be used as feedback to a controller to control the intensity of the administered stimulus.

Various methods are used to measure affective states like measuring the concentration of various hormones in the blood stream, judging affective state from facial expressions, body movements and gestures etc. While affective state measurement from bloodwork is very accurate, it is intrusive, requires clinical practitioners and is not real-time. A trained human observer can provide real time affective state assessment based on facial expressions and body movements but that requires specialized human labor. Also, they are susceptible to human bias, will be inconsistent in between observers and is not scalable. Computer vision techniques have been used to automate the process of detecting affective states from facial expressions [7] but those techniques can suffer from bias in the training dataset due to the fact that individual subjects have different expressions corresponding to different affective states and they might choose to deliberately mask their expressions if they do not want to reveal their emotions.

On the other hand, physiology depends on the autonomic nervous system which controls the involuntary functions of the body and cannot be voluntarily controlled or suppressed. Also, physiology is largely consistent among individuals and hence provide a reliable and consistent measurement of affective states.

In this paper, relevant literature on affective state detection using physiology data is discussed. An experiment to classify affective states of meditation, amusement and high cognitive load using functional near infrared spectroscopy is described and the results are discussed in detail. A mean accuracy of 83.04% was achieved in three class classification with an individual model; 84.39% accuracy was achieved for a group model and 60.57% accuracy was achieved for subject independent model using leave one out cross validation.

## II. RELATED WORK

Autonomic physiological responses are highly correlated to affective states, and these responses can be measured by various non-invasive sensors. The most used sensors are Electrocardiogram (ECG), Electroencephalography (EEG), PPG or blood volume pulse (BVP), Electrodermal Activity (EDA) or Galvanic Skin Response (GSR). These signals are commonly chosen due to their strong correlation to affective states like stress, which is one of the most significant and widely researched affective state. Sensors like thermal imaging and accelerometers are also used.

Table 1 shows some of the existing work done on measurement of affective states using physiology data and represents the state of the art in this field. The authors in [8] created a public dataset WESAD which has been featured in various other works. It is a multimodal dataset containing BVP, EDA, temperature and accelerometer sensor data corresponding to stressed and not stressed affective states. It is a labelled dataset with 15 participants. The authors provided a benchmark of 88.25% accuracy using the leave one subject out validation method and similar results have been replicated by some other studies as well. While attempting to replicate the results, it was observed that baseline vs. stress classification was trivial and not indicative of stress detection. If the model is fed data corresponding to any other affective state, the prediction will still be that of stress. This is because individual participant data is normalized with respect to the baseline and then the problem is reduced to classification of values that have very distinct proximity to zero. Similar methods have been used by the authors in [9], achieving a 96.5% classification accuracy for a subject independent model. The studies described in [12,13] used similar methods using thermal imaging and heart rate variability sensors and achieved above 80% subject independent classification accuracy. The authors in [10] have created a group model for stressed/ not stressed binary classification with 250 participants and achieved 81.5% classification accuracy with neural networks using 5-fold cross validation method. Li et al. [11] achieved above 99.5% accuracy in a baseline/ stress/ amusement classification task. Such high accuracies were possible due to a 'look ahead' bias, where the entire sample set was shuffled before being divided into training and testing sets. This is a common practice in many machine-learning applications but may not be an appropriate method for dealing with physiology signals since physiology signals change slowly and continuously, and when sampled at a high enough frequency, the values of successive samples are very similar. Hence if the data is shuffled prior to dividing it into training and testing sets, the training set contains values that are very similar to that of the testing set, and hence the model effectively is exposed to the testing set data during the training process.

TABLE I: Related Work

| Authors | Number of subjects | Labels | Modalities | Classifiers | Validation method | Prediction rate |
|---|---|---|---|---|---|---|
| Schmidt et al. [8] | 15 | Stress/ not stressed | BVP, EDA, Temp, Acc | Random forest | LOSO | 88.25% |
| Youngjun et al. [9] | 17 | Stress/ not stressed | BVP, Temp | Neural network | LOSO | 96.5% |
| Taylor et al. [10] | 250 | Stress/ not stressed | SC, temp, accelerometer | Neural Network | 5-fold cross validation | 81.5% |
| Li et al. [11] | 15 | Baseline/ stress/ amusement | BVP, EDA, Temp, Acc | CNN, MLP | 10-fold cross validation | 99.8%, 99.55% |
| Cardone et al. [12] | 10 | Stress/ not stressed | Thermal Imaging | SVR | LOSO | 0.80 AUC |
| Castaldo et al. [13] | 42 | Stress/ not stressed | Heart rate variability | SVM | LOSO | 88% |

## III. OBJECTIVE OF THE STUDY

In the related works section, most of the prior work discussed has performed a binary classification of stress/ not stressed and one has performed a three-class classification of stressed/ amused/ baseline. Humans are capable of a large number of affective states which have varying degree of proximity to each other in the affective state spectrum and using a binary model to predict a certain affective state is unreliable. This is due to the fact that if the model is trained on only two affective states or one affective state and baseline, any other affective state that the model is not trained on will be wrongly classified. Also, careful attention must be paid to ensure that the measured values are caused by the target affective state only, and not by some other activity that the participant might be performing simultaneously. To address these concerns, ideally an experiment to evoke all affective states individually and exclusively should be performed, which in reality is very difficult because the same stimulus may evoke different affective states in different participants based on their individual preferences and life experiences. In this study, the experiment will aim to evoke three affective states: meditation, amusement, and high cognitive load. These three affective states were initially chosen because these can be evoked reliably on various participants and will provide a good starting point from which the study can be expanded to include other affective states. The fNIRs is a relatively new sensor and not many open-source datasets for affective state detection using fNIRs is available. This study will also serve to validate the usability and reliability of this sensor for this purpose.

## IV. NEAR INFRARED SPECTROSCOPY

There are several physiological sensors in use such as functional magnetic resonance imaging (FMRI), EEG, ECG, heart rate variability, skin conductance etc. that have been used in many studies. Each of these sensors have their own advantages and disadvantages. FMRI has excellent spatial resolution and is resilient to environmental factors, but the apparatus required is bulky and expensive and is not portable, hence is of limited practical use for the scope of applications discussed in this paper. EEG, ECG and accelerometers have very high temporal resolution which makes them suitable for activity recognition because they can detect short duration motions. But they are also susceptible to contamination by high frequency noise and motion artefacts which require complicated filtering in software. They are also influenced by ambient temperature, activities of the participants and hence have difficulty separating the effects of affective state vs activities of the participants.

Over the last decade and half, optical brain imaging using functional near infrared spectroscopy (fNIRs) have gained traction and have been used in various settings. The key feature of fNIRs sensor is that it measures the concentration changes of oxygenated and deoxygenated hemoglobin in the different regions of the brain, which is a relatively slow process, hence the signal is not contaminated by any motion artefacts. The sensor consists of a series of infrared LEDs of one or more wavelengths and a series of photodiodes. The LEDs emit infrared light which travels into the scull cavity and the reflected light is received by the photodiodes. The wavelengths are selected such that at those wavelengths, skin and bone are transparent to light. Hence the only components to reflect light are water, oxygenated and deoxygenated hemoglobin. Fig. 1 shows the absorption spectrum of these three components. From the figure, it can be seen that at wavelengths between 700nm to 900nm, only oxygenated and deoxygenated hemoglobin has significant response to light, with their response being almost identical at 790nm. This is called the isosbestic point. Typically, a wavelength is selected on either side of this isosbestic point to selectively measure the response from the oxygenated and deoxygenated hemoglobin in blood.

The optical path followed by the light from the emitter to the detector depends on the beam angle of the LED and the distance between the LEDs and the photo diodes. Fig. 2 shows an LED with two detectors at two different distances. It can be seen that the photodiode located closer to the LED receives light through a shallower optical path than the photo diode that is located further away. Shallower optical paths lead to measurement of oxygenated and deoxygenated hemoglobin concentration changes in the superficial capillaries in the skin and the deeper optical pathways measure those changes in the tissues of the brain. Hence, by measuring the short optical pathways, called 'short channel' signals, concentration changes due to activities like blinking, eye-brow movements can be detected and eliminated from the long channel signals to further reduce motion artefacts.

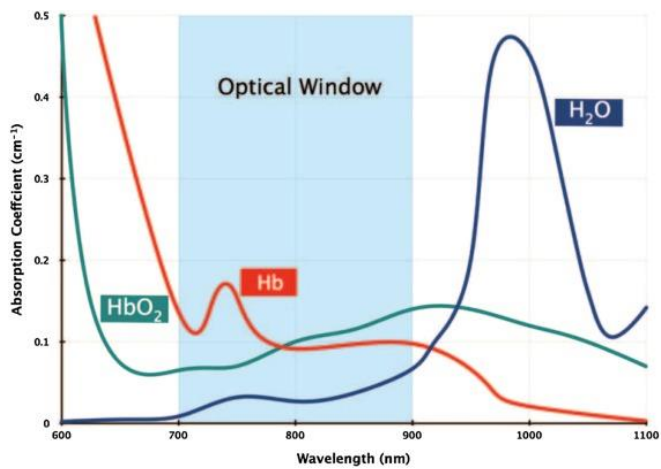

Fig. 1. Absorption spectrum of water, oxygenated and deoxygenated blood

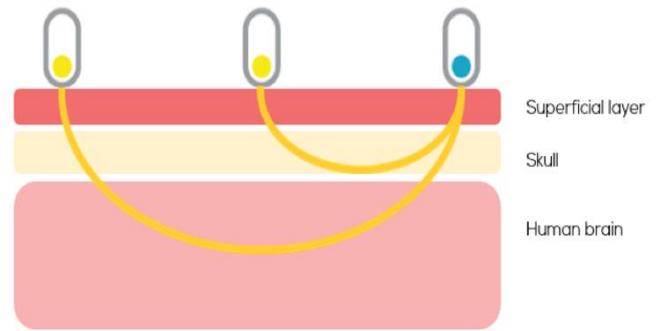

Fig. 2. Optical pathway of infrared light from emitter to detectors

## V. DATA PREPROCESSING

The data collected by fNIRs consist of three signals per channel: reflection of light of wavelength 730nm, reflection of light of wavelength 850nm and reading when LEDs of both wavelengths are turned off. While the data is relatively noise free, it does contain a few other signals that are not of importance to our application. These signals include regular fluctuations due to heartbeat (1-1.5 Hz), respiration (0.2-0.3 Hz) and Mayer's wave (0.1 Hz), which are the cyclic waves in arterial blood pressure brought about by oscillations in baroreceptor and chemoreceptor reflex control systems. These are eliminated by filtering the signal using a 0.1Hz low pass filter.

The photodiodes also receive small amounts of ambient light which result in some current. This current is measured while the LEDs are turned off, called the dark current. The dark current represents the sum of the bias current in the photodiodes and current due to ambient light. This value is subtracted from the signal to eliminate any device and ambience specific bias.

Small facial movements like moving eyebrows and blinking may have some effect on the blood concentration in the superficial capillaries which also influence the readings of the sensor. These are eliminated by taking the short channel data and subtracting it from the long channel data.

This concludes the denoising of the signal. Then the baseline data collected in the first few minutes of the experiment is subtracted from the signal to account for individual differences in participants. Then the signal is decomposed into four features per channel: oxygenated hemoglobin concentration (HbO) from 850nm data, deoxygenated hemoglobin concentration (HbR) from 730nm data, total hemoglobin concentration (HbT = HbO + HbR) and oxygen consumption (Oxy = HbO – HbR).

The device used in this study is a 16 channel fNIR device manufactured by fNIRDevices (www.fnirdevices.com) which uses 4 dual wavelength LEDs (730nm and 850nm) and 10 photodiodes. The optodes or optical imaging regions are located in between each photodiode and LED pair. Fig. 3 shows the optode layout of a typical fNIR device. This results in 16 optodes that provide the 16 channels of data. This device also contains two reference channels that are located close to the LEDs which provide the short channel measurement. The LEDs are switched on at different wavelengths and turned off alternately and the measurements are taken, which results in a total of 48 channel of data (730nm, 850nm and ambient times

16). After the preprocessing steps mentioned above, the final feature set includes 16 * (HbO, HbR, HbT and Oxy) resulting in 64 features per sample. Fig. 4 shows a raw signal trace and fig. 5 shows a processed signal trace.

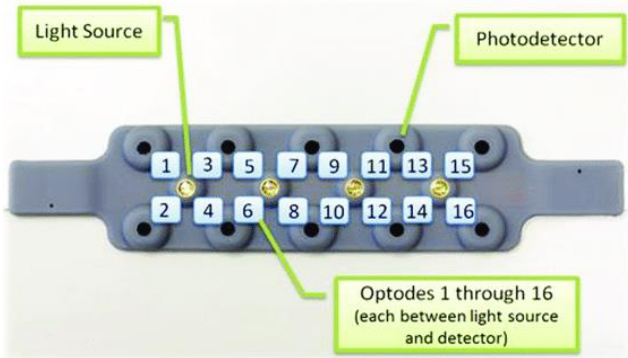

Fig. 3. Optode layout of a typical fNIR device

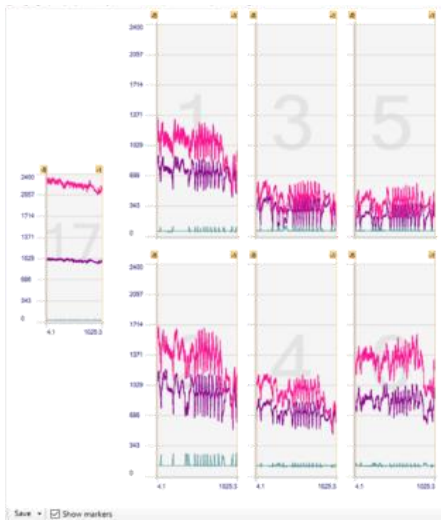

Fig. 4. Raw fNIR signal, pink denotes oxygenated hemoglobin concentration, violet denotes deoxygenated hemoglobin concentration and green denotes dark current

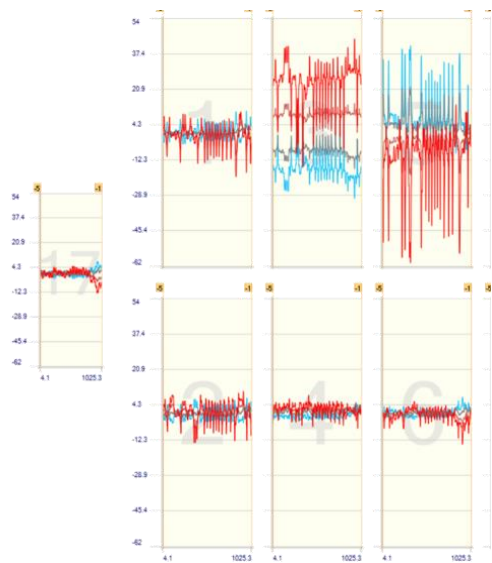

Fig. 5. Processed fNIR signal, orange represents HbO, brown represents HbR, red represents HbT and blue represents Oxy

## VI. EXPERIMENT

An experiment was conducted with the aim of eliciting different affective states in a controlled environment in order to test the accuracy of classification of the affective states using fNIR data. 10 participants were recruited for the experiment, data for 1 participant had to be discarded due to faulty sensor readings. Among them were 5 males (mean age 25.4 std. 2.05) and 4 females (mean age 25.5 std. 3.35). The experiment protocol was as follows: 2 minutes of baseline data collection during which the participants were instructed to sit idle. This data is important to account for individual differences as well as the effects of prior activity, caffeine intake etc. This was followed by three four-minute sessions of meditation while listening to a guided meditation music, four minutes of watching stand-up comedy by a comedian of their choice and four minutes of continuously writing the Fibonacci sequence with a pen and paper. The data was collected at a sampling frequency of 4 Hz. The sessions were timed with one minute of dead band in between to allow the participant to fully switch affective states before data collection in order to prevent false labelling. Fig. 6 shows a participant wearing the fNIR headset.

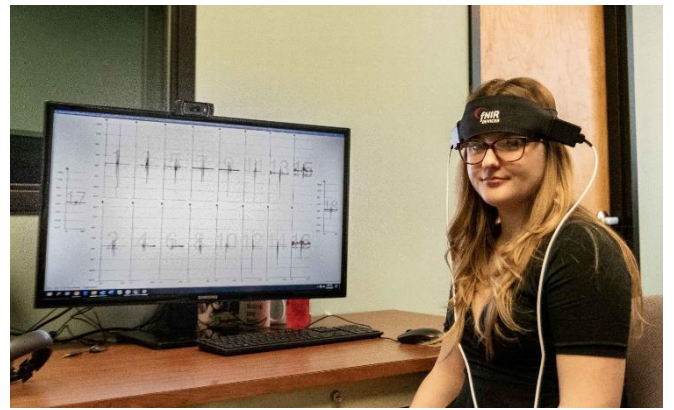

Fig. 6. A participant wearing the fNIR headset before an experiment

## VII. RESULTS

Three intended use cases were identified for the machine learning models:

1) *The model will be used to predict the affective state of one particular individual, whose historic data can be accessed*

2) *The model will be used to predict the affective state of a group of individuals, whose historic data can be accessed*

3) *The model will be used to predict the affective state of unknown individuals, whose historic data can not be accessed*

The first two use cases can be applied in workplaces or nursing homes or any such settings where the same individual/ group of individuals will use the system over a period of time. The third use case is important for deployment, where no knowledge of the end user is available.

To address these use cases, three different machine learning models were trained:

Case 1:
The individual model was trained on 80% of an individual's data and tested on the remaining 20% of the individual's data. This was repeated 100 times for each individual. During each run, a contiguous 20% of the data from each of the classes were set aside for testing, and then the training set and testing set were shuffled. It was ensured that each class had the same number of samples to keep the training data balanced, and shuffling was performed after dividing the dataset into training and testing sets to prevent the look-ahead bias discussed at the end of section II.

Random forest and ANN's performed the best, with ANNs slightly outperforming the random forest. A neural network with 64 input nodes, 2 hidden layers with 128 nodes each and an output layer with 3 nodes was used for the individual model. This model was intentionally overfitted to some extent because it is supposed to have access to each individual's prior data and does not need to generalize between participants. The hidden layers had 'relu' as activation function and the final layer used softmax. The 'Adam' optimizer was used from the Keras library in python with sparse categorical cross entropy as the loss function. The best results were obtained using training parameters of batch size 5 and 8 epochs.

Table 2 shows the mean, max, min and standard deviation of accuracies of each individual participant and fig. 7 shows the error diagram.

Table II. Individual model performance

| Metric | Mean | Std. | Max | Min |
|---|---|---|---|---|
| Participant 1 | 79.96% | 7.30% | 93.16% | 69.99% |
| Participant 2 | 78.99% | 8.50% | 92.00% | 68.33% |
| Participant 3 | 67.62% | 6.54% | 80.33% | 60.66% |
| Participant 4 | 70.60% | 10.75% | 99.00% | 56.99% |
| Participant 5 | 96.22% | 9.94% | 100% | 66.66% |
| Participant 6 | 83.25% | 7.95% | 95.50% | 70.17% |
| Participant 7 | 80.90% | 7.77% | 89.00% | 67.50% |
| Participant 8 | 90.91% | 10.57% | 100% | 66.67% |
| Participant 9 | 98.93% | 1.40% | 100% | 95.66% |

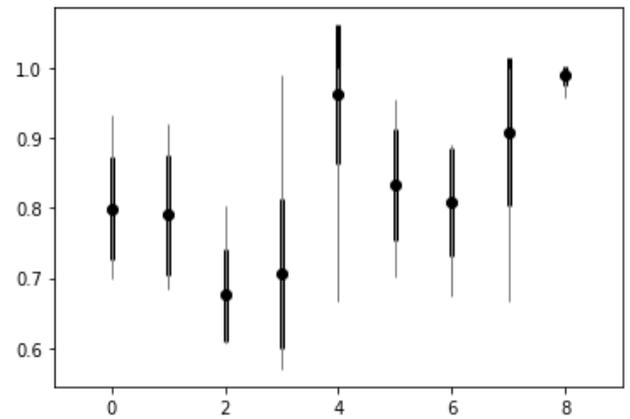

Fig. 7 Individual model error diagram

Case 2:

The group model was trained on 80% of all participants' data combined and tested on the remaining 20% of all participants' data. Again, the data was divided into training and testing sets prior to shuffling to prevent the look-ahead bias. For the group model, a neural network with 64 input nodes, 2 hidden layers with 256 nodes each and an output layer with 3 nodes was used. The model used the same activation function and loss as the previous model. The training parameters were batch size of 40 and 5 epochs. The table 3 shows the performance of the group model and fig. 8 shows the confusion matrix.

Table III. Group model performance

| Metric | Value |
|---|---|
| Mean | 84.39% |
| Std. | 2.30% |
| Max | 87.18% |
| Min | 81.22% |

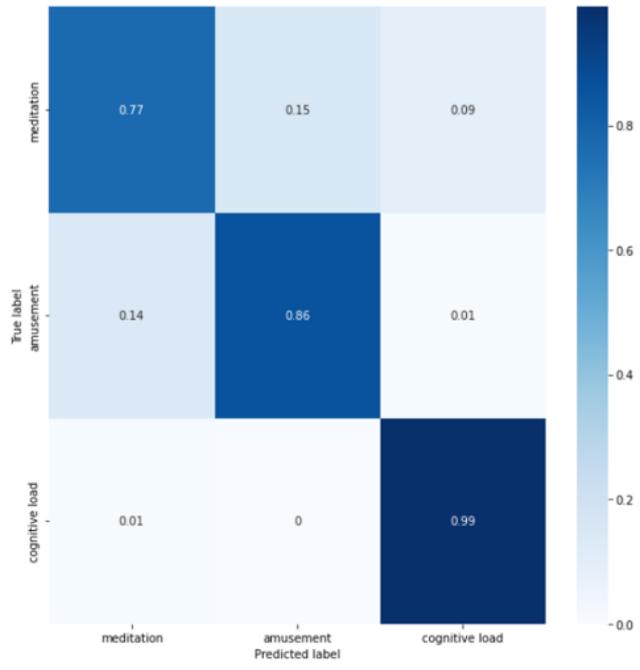

Fig. 8 Group model confusion matrix

From the confusion matrix of the group model, it can be clearly seen that the model could classify cognitive load with a very high degree of certainty but there was some confusion between meditation and amusement. This has been addressed in the next section.

Case 3:

The subject independent model was trained on 100% data of all but one participant and tested on 100% data of the remaining participant, often called leave one subject out validation (LOSO). This was repeated 100 times for each participant and the results were averaged. For the subject independent model, a neural network with 64 input nodes, 2 hidden layers with 256 nodes each and an output layer with 3 nodes was used. The model used the same parameters as the previous models. The training parameters were batch size of 120 and 5 epochs. The table 4 shows the performance of the subject independent model.

Table IV. Subject independent model performance

| Metric | Group |
| --- | --- |
| Mean | 60.57% |
| Std. | 12.82% |
| Max | 83.99% |
| Min | 39.17% |

## VIII. Interpretation of the Confusion Matrix

From the confusion matrix in the previous section, it is evident that the model could classify cognitive load with a high degree of confidence, but it was not the case for the other two classes. There can several possible explanations for this observation:

1) In the affective state spectrum, meditaion and amusement are much closer to each other than high cognitive load, hence it was easier for the model to clearly draw a decision boundary between cognitive load and the other two states.

2) Meditation is an acquired skill, and it takes a lot of regular practice to achieve an uninterrupted meditative state. It is possible that the participants' mind veered off during the meditation period to other thoughts, some of which may have been amusing, in which case it would have contaminated some of the data.

3) Another possible explanation is that while the meditation and amusement states were evoked using computer based stimulus where no motor skills were involved, the cognitive load task was pen and paper based and involved the participant writing the numbers in addition to mental calculation. It is possible that voluntary motor control contributed to easier identification of this task. An experiment was designed to verify that cognitive load could be detected independent from voluntary motor control.

## IX. Experiment to Verify Effect of Motor Control

A second experiment was designed to study the effect of voluntary motor control and verify if high cognitive load could be identified independent of motor control. The experiment was conducted with four participants and the protocol was as follows:

2 minutes of baseline when the participant was sitting idle, 3 minutes of Fibonacci sequence addition and 3 minutes of writing their own name.

Both of these tasks required the same degree of voluntary motor control but in the case of Fibonacci sequence addition, it required high level of cognition while during writing their own name, minimal cognitive capabilities were employed. The data was preprocessed using the same steps as outlined in section V. A group model was trained to perform binary classification between these two activities, with 80% of all data from all four participants combined as the training set and the remaining 20% data as the testing set. A neural network was used for the classification and an accuracy of 81.58% was achieved with a standard deviation of 0.4%. This experiment clearly demonstrated that it was possible to distinguish cognitive overload independently from voluntary motor control.

## X. Discussion

In this study, we created three models to address three different real-life scenarios. The individual model can be trained for each individual who is going to use the system, such a model is practical in an institutional setting where the same individual will use the system multiple times over a period of time. The group model is more relevant where the same equipment will be shared among a number of individuals multiple times over a period of time, but individual profiles may not be feasible, e.g., shared infrastructure in long term care facilities often have record of all individuals who use are enrolled to use an equipment but do not always record who is using it at a given time. We are developing this system as a feedback controller to regulate the level of difficulty of activities designed for older adults living in long term care facilities, and in such use cases, the group model would be particularly useful. The subject independent model is designed for use by individuals whose historic data will not be accessible for training, it will be useful in deployment as a product to the end user. In real life, individuals exhibit a spectrum of affective states, hence binary classification is of little use. We attempted to create a three-class classification and achieved an individual model accuracy of 83.04% which is almost at per with the state of the art. The group model accuracy was also above 80%. The subject independent model accuracy was just above 60% which can be improved with hyper-parameter tuning. The group model accuracy is less than some findings in other literature, but that is because most have done a two-class classification or three class classification where one class is baseline. Also, some authors have achieved very high accuracies due to shuffling the data before dividing it into training and testing sets which leads to a look-ahead bias. We also verified that the greater confidence in classifying the cognitive overload was not due to voluntary motor control during that task. This numbers are a realistic expectation of what we might see in real life deployment. We will continue this work to accommodate more affective states and try to improve the subject independent model accuracy. A robust model will enable a better feedback system to smart controllers and enable adaptive human computer/ human machine interactions to take place.

## REFERENCES

[1] Changchun Liu, Karla Conn, Nilanjan Sarkar, Wendy Stone, Physiology-based affect recognition for computer-assisted intervention of children with Autism Spectrum Disorder, International Journal of Human-Computer Studies, Volume 66, Issue 9, 2008, Pages 662-677, ISSN 1071-5819

[2] Sano A, Taylor S, McHill A, Phillips A, Barger L, Klerman E, Picard R, Identifying Objective Physiological Markers and Modifiable Behaviors for Self-Reported Stress and Mental Health Status Using Wearable Sensors and Mobile Phones: Observational Study, J Med Internet Res 2018;20(6):e210.

[3] S. Wu and T. Lin, "Exploring the use of physiology in adaptive game design," *2011 International Conference on Consumer Electronics, Communications and Networks (CECNet)*, 2011, pp. 1280-1283, doi: 10.1109/CECNET.2011.5768186.

[4] N. Oliver and L. Kreger-Stickles, "Enhancing Exercise Performance through Real-time Physiological Monitoring and Music: A User Study," *2006 Pervasive Health Conference and Workshops*, 2006, pp. 1-10, doi: 10.1109/PCTHEALTH.2006.361660.

[5] Trejo, Leonard J., et al. "Experimental design and testing of a multimodal cognitive overload classifier." Foundations of Augmented Cognition 2007 (2007): 13-22.

[6] Morton, Jessica, et al. "Identifying predictive EEG features for cognitive overload detection in assembly workers in Industry 4.0." H-Workload 2019: 3rd International Symposium on Human Mental Workload: Models and Applications (Works in Progress). 2019.

[7] Haines, Nathaniel & Southward, Matt & Cheavens, Jennifer & Beauchaine, Theodore & Ahn, Woo-Young. (2019). Using computer-vision and machine learning to automate facial coding of positive and negative affect intensity. PLOS ONE. 14. e0211735. 10.1371/journal.pone.0211735.

[8] P. Schmidt, A. Reiss, R. Duerichen, and K. V. Laerhoven, "Introducing wesad, a multimodal dataset for wearable stress and affect detection,"ICMI 2018 - Proceedings of the 2018 International Conference onMultimodal Interaction, pp. 400–408, 2018.

[9] S. J, B.-B. N. C. Youngjun, and Julier, "Instant stress: Detectionof perceived mental stress through smartphone photoplethysmographyand thermal imaging,"JMIR Ment Health, vol. 6, p. e10140, 4 2019.[Online]. Available: http://www.ncbi.nlm.nih.gov/pubmed/30964440

[10] S. Taylor, N. Jaques, E. Nosakhare, A. Sano, and R. Picard, "Person-alized multitask learning for predicting tomorrow's mood, stress, andhealth,"IEEE Transactions on Affective Computing, vol. 11, pp. 200–213, 2020.

[11] R. Li and Z. Liu, "Stress detection using deep neural networks,"BMCMedical Informatics and Decision Making, vol. 20, p. 285, 2020.[Online]. Available: https://doi.org/10.1186/s12911-020-01299-4

[12] D. Cardone, D. Perpetuini, C. Filippini, E. Spadolini, L. Mancini,A. M. Chiarelli, and A. Merla, "Driver stress state evaluation by meansof thermal imaging: A supervised machine learning approach basedon ecg signal,"Applied Sciences, vol. 10, 2020. [Online]. Available:https://www.mdpi.com/2076-3417/10/16/5673

[13] R. Castaldo, L. Montesinos, P. Melillo, C. James, and L. Pecchia,"Ultra-short term hrv features as surrogates of short term hrv: a casestudy on mental stress detection in real life,"BMC Medical Informaticsand Decision Making, vol. 19, p. 12, 2019. [Online]. Available:https://doi.org/10.1186/s12911-019-0742-y